\documentclass[conference]{IEEEtran}
\IEEEoverridecommandlockouts

\usepackage{graphicx}
\usepackage{amsmath,amssymb}
\usepackage{cite}
\usepackage{booktabs}
\usepackage{multirow}
\usepackage{xcolor}
\usepackage{tikz}
\usepackage{pgfplots}
\usepackage{caption}   
\usepackage{hyperref}  
\usepackage{algpseudocode}
\usepackage[ruled,vlined,linesnumbered]{algorithm2e}
\usepackage{url}
\usepackage{pgfplots}
\usepgfplotslibrary{groupplots}
\pgfplotsset{compat=1.18}
\usepackage{listings}
\usepackage{xcolor}
\usepackage{pifont}
\usetikzlibrary{arrows.meta,shapes.geometric,positioning}
\newcommand{\xmark}{\ding{55}}%
\usepackage{pgfplotstable}
\usepackage{siunitx}
\usepackage{subcaption} 

\title{\emph{AFarePart}: \underline{A}ccuracy-aware \underline{Fa}ult-\underline{re}silient \underline{Part}itioner for DNN Edge Accelerators}

\author{
\IEEEauthorblockN{
Mukta Debnath\IEEEauthorrefmark{1},
Krishnendu Guha\IEEEauthorrefmark{2},
Debasri Saha\IEEEauthorrefmark{1},
Amlan Chakrabarti\IEEEauthorrefmark{1},
Susmita Sur-Kolay\IEEEauthorrefmark{3}
}
\IEEEauthorblockA{
\IEEEauthorrefmark{1}\textit{University of Calcutta, India} \quad
\IEEEauthorrefmark{2}\textit{University College Cork, Ireland} \quad
\IEEEauthorrefmark{3}\textit{Indian Statistical Institute, India}
}
\IEEEauthorblockA{
Email: muktapearl.debnath@gmail.com,
kguha@ucc.ie,
debasri\_cu@yahoo.in,\\
acakcs@caluniv.ac.in,
ssk@isical.ac.in
}
}

\begin{document}

\maketitle

\begin{abstract}
Deep Neural Networks (DNNs) are increasingly deployed across distributed and resource-constrained platforms, such as System-on-Chip (SoC) accelerators and edge-cloud systems. DNNs are often partitioned and executed across heterogeneous processing units to optimize latency and energy. However, the reliability of these partitioned models under hardware faults and communication errors remains a critical yet underexplored topic, especially in safety-critical applications. In this paper, we propose an accuracy-aware, fault-resilient DNN partitioning framework targeting multi-objective optimization using NSGA-II, where accuracy degradation under fault conditions is introduced as a core metric alongside energy and latency. Our framework performs runtime fault injection during optimization and utilizes a feedback loop to prioritize fault-tolerant partitioning. We evaluate our approach on benchmark CNNs including AlexNet, SqueezeNet and ResNet18 on hardware accelerators, and demonstrate up to 27.7\% improvement in fault tolerance with minimal increase in performance overhead. Our results highlight the importance of incorporating resilience into DNN partitioning, and thereby paving the way for robust AI inference in error-prone environments.


\end{abstract}


\begin{table*}[ht]
\renewcommand{\arraystretch}{1.2}
\centering
\caption{Comparison of DNN Partitioning and Scheduling Tools}
\label{tab:tool_comparison}
\begin{tabular}{|l|l|c|c|l|}
\hline
\textbf{Tool} & \textbf{Partitioning Strategy} & \textbf{Fault Aware} & \textbf{Custom HW} & \textbf{Objectives} \\
\hline
\textbf{CNNParted \cite{cnnparted}} & NSGA-II Pareto partitioning & \xmark & Edge Accelerators & Latency, Energy\\
\hline
\textbf{Edgent \cite{edgent}} & RL-based adaptive partitioning & \xmark & Edge–Cloud & Latency\\
\hline
\textbf{Neurosurgeon \cite{neurosurgeon}} & Profiling-based layer placement & \xmark & Mobile–Cloud & Latency, Energy \\
\hline
\textbf{SPINN \cite{SPINN}} & HW-aware scheduling and codegen & \xmark & Custom accelerators & Latency, Energy \\
\hline
\textbf{RoadRunner \cite{roadrunner}} & Auto DNN partitioning + runtime scheduling & \xmark & FPGA + GPU & Latency, Energy \\
\hline
\textbf{MAESTRO \cite{MAESTRO}} & Cost model + mapping optimizer & \xmark & Analytical only &  Energy \\
\hline
\textbf{This Work (Ours)} & Fault-aware Pareto + dynamic reconfig. & \checkmark & Edge Accelerators & \textit{\textbf{Accuracy}}, Energy, Latency \\
\hline
\end{tabular}
\end{table*}

\section{Introduction}

Deep Neural Networks (DNNs) have revolutionized computer vision and are widely deployed in safety-critical applications such as autonomous driving, medical diagnostics, and industrial automation. These applications often demand not only high accuracy but also stringent constraints on latency, energy efficiency, and system reliability. In order to meet these requirements, DNN inference is increasingly executed in a distributed manner across heterogeneous hardware accelerators, such as FPGAs, CPUs, and NPUs on System-on-Chip (SoC) platforms. For reduction of computational burden and meeting deployment constraints, DNNs are typically partitioned into sub-networks executed across different nodes or accelerators~\cite{DNNEmbed}. Tools such as CNNParted~\cite{cnnparted}, RoaD-RuNNer~\cite{roadrunner} have demonstrated the effectiveness of such partitioning in optimizing latency and energy consumption. However, these tools assume ideal hardware conditions and do not consider the impact of faults, which are increasingly prevalent due to hardware aging, aggressive voltage scaling, or radiation-induced soft errors. DNN inference on accelerators like Eyeriss~\cite{chen2019eyeriss} and SIMBA~\cite{symba} is also vulnerable to soft errors, aging-induced degradation, and voltage-scaling-induced timing faults, all of which can cause bit-flips in activations or weights, and thus leading to incorrect predictions~\cite{sifi}. 

In real-world deployments, hardware faults (e.g., bit-flips from voltage glitches or electromagnetic interference) and communication errors (e.g., packet loss) can severely degrade DNN inference. In partitioned networks, these types of faults may propagate across partitions, amplifying their impact and compromising accuracy and reliability. Under transient faults, a generated partitioning solution may be non-reproducible for the same problem instance, preventing users from obtaining consistent, high-quality mappings. Given the prevalence of faulty hardware and unstable electrical environments, partitioning tools must incorporate mechanisms to produce solutions with minimal fault susceptibility. However, most existing strategies remain fault-agnostic, leaving fault-resilient partitioned DNN  hardware accelerators largely unexplored.


In order to address this gap, we propose \emph{AFarePart} — an Accuracy-aware Fault-resilient Partitioning Framework for DNN inference — designed to mitigate both naturally occurring soft errors and adversarially induced transient faults in fault-prone edge hardware accelerators. Our framework incorporates fault injection during the partition evaluation process, feeding fault-induced accuracy degradation back into the multi-objective optimizer that enables real-time evaluation of accuracy under injected faults. By treating accuracy degradation as a first-class optimization objective, alongside latency and energy, our system produces robust partitioning strategies that maintain high inference quality, even under adverse fault conditions. We offer the following key contributions:
\begin{itemize}
    \item a partitioning methodology that incorporates fault injection outcomes, specifically from corrupted activations and weights, into the optimization process;
    \item introducing accuracy degradation under fault as a first-class optimization objective, alongside energy and latency, within a Pareto-based evolutionary loop;
    \item evaluating our fault-resilient framework on different CNNs like AlexNet, SqueezeNet, and ResNet18, across different fault rates and hardware profiles; and
    \item assessing the effectiveness of our proposed method, demonstrating up to a 27.7\% improvement in fault resilience with only a minimal increase in latency and energy overhead.
\end{itemize}

   In the rest of this article,  Section~\ref{sec:background} discusses related works and the limitations of current partitioning strategies. Section~\ref{sec:fault-model} describes our assumed fault model and the threat surface. Section~\ref{sec:prob-def} defines the problem. Section~\ref{sec:framework} introduces our fault-aware partitioning methodology. Section~\ref{sec:results} presents the evaluation setup and results. Section~\ref{sec:conclusion} concludes with possible future directions.

\section{Related Works}
\label{sec:background}
The increasing deployment of Deep Neural Networks (DNNs) across edge and embedded platforms has led to a growing interest in efficiently partitioning frameworks that can balance latency, energy, and resource constraints. In this context, several tools have been developed with varying objectives and hardware support. However, most existing solutions either neglect fault tolerance or lack runtime adaptability in adverse operating conditions.~\autoref{tab:tool_comparison} presents a comprehensive comparison of such approaches, derived from our literature survey, with our proposed framework.

\textit{CNNParted}~\cite{cnnparted} is a recent tool that partitions CNNs for distributed inference across heterogeneous hardware by optimizing for latency and energy using multi-objective evolutionary algorithms. However, CNNParted operates under fault-free assumptions and does not incorporate any reliability or robustness metric in its optimization strategy. \textit{RoaDRuNNer}~\cite{roadrunner} is a real-time DNN execution scheduler that focuses on performance predictability and task scheduling on edge SoCs. It supports runtime offloading of tasks based on deadline violations and processor slack time. Despite its dynamic adaptability, RoadRuNNer does not account for hardware faults or model accuracy under fault-injected scenarios. \textit{SPINN}~\cite{SPINN} and \textit{GuardianNN}~\cite{GuardianNN} explore security-aware and resilient AI execution by considering adversarial attacks and hardware threats. However, they do not implement fault-aware accuracy as a quantitative optimization objective and lack fine-grained resource-aware partitioning.
Other partitioning tools, such as \textit{SplitAI}~\cite{splitai} and \textit{Neurosurgeon}~\cite{neurosurgeon}, focus on partitioning DNNs between edge devices and the cloud based on latency and energy profiles but similarly omit fault tolerance as an optimization objective. \textit{HeteroSplit}~\cite{heterosplit} extends this line of work for hardware heterogeneity but lacks support for fault injection or resilience analysis.

Fault injection techniques have been studied independently to assess the robustness of DNNs, particularly in safety-critical applications~\cite{fi-cnn1, fi-cnn2}. 
Liu et al.~\cite{FIDNN} investigate the vulnerability of DNNs to fault injection attacks aimed at inducing targeted misclassification. They propose two attack strategies: Single Bias Attack (SBA), which modifies a single parameter in the DNN to force misclassification, and Gradient Descent Attack (GDA), which minimizes collateral impact by carefully perturbing parameters using a gradient-guided approach. Hoefer et al. introduce \textit{SiFi-AI}~\cite{sifi}, a flexible fault injection framework designed to evaluate the vulnerability of AI accelerators under soft errors. SiFi-AI supports various fault models, including bit-flip injections in both memory and computational logic, enabling detailed analysis of fault propagation and its impact on deep learning inference accuracy. These studies show that faults in weights or activations can cause significant accuracy degradation, motivating the need for resilience-aware optimization. However, integration of such analysis into the partitioning process has not been studied.

\begin{figure}[ht]
    \centering
    \includegraphics[width=0.7\linewidth]{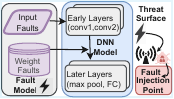} 
    \caption{Fault Model and Threat Surface}
    \label{fig:fault-model}
\end{figure}

\section{Fault Model and Threat Surface}
\label{sec:fault-model}
In order to evaluate the fault resilience of partitioned DNNs deployed on heterogeneous SoC accelerators, we adopt a fault model that encompasses both naturally occurring soft errors and adversarially induced transient faults~\cite{deepstrike, Bit-Flip-Crushing, DNNGuard, HTDNNSurvey, FINN, sifi}. This unified model supports systematic injection of faults during optimization, enabling resilience-aware partitioning across various hardware accelerators platforms such as \textit{Eyeriss}~\cite{chen2019eyeriss} and \textit{SIMBA}~\cite{symba}. Our assumed fault mode and threat scenarios are illustrated in~\autoref{fig:fault-model}.

\subsection{Threat Model and Attack Surface}
Attackers can induce transient faults during inference to manipulate DNN predictions and cause targeted misclassifications. In distributed DNN execution (e.g., early layers on Eyeriss and later layers on SIMBA), attackers may selectively target platform-specific regions with high impact. We consider two primary adversarial fault vectors:
\noindent
\begin{itemize}
  \item Voltage Glitching: Deliberate under/over-volting to flip bits in memory or compute units, often affecting buffers or MAC logic~\cite{deepstrike}.
  \item Electromagnetic (EM) Injection: Use of directed pulses or laser to flip specific bits in memory arrays, often targeting critical DNN layers~\cite{HTDNNSurvey}.
\end{itemize}

These attacks can originate from insider threats, malicious vendors, or physical access adversaries, with the goal of silently degrading accuracy or causing functional misclassifications in safety-critical applications.

\subsection{Fault Assumptions and Scope}

Our fault model, captures realistic hardware-induced fault scenarios in quantized neural networks with edge-deployed DNN accelerators.
We adopt a bit-flip fault injection strategy focused on the least significant bits (LSBs) of quantized weights to model such faults in quantized neural networks. As modern hardware accelerators for neural networks like Eyeriss and SIMBA typically operate on fixed-point integer representations (e.g., INT8), injecting faults post-quantization accurately captures the nature of soft errors that occur in hardware due to voltage noise, timing violations, or radiation-induced bit flips. LSB faults are particularly common in low-power embedded systems due to reduced error correction circuitry and aggressive voltage scaling, and while their impact on computation is lower compared to most significant bit (MSB) faults, they occur more frequently and cumulatively degrade performance.

\noindent

We focus exclusively on transient soft errors — single-event, non-destructive bit-flips that manifest during inference due to environmental or adversarial conditions. Permanent hardware defects (e.g., stuck-at faults or hardware Trojans~\cite{HTDNNSurvey}) are out of scope of this work, as these require structural mitigation. We simulate faults in two key domains: \textit{activation faults} and \textit{weight faults}. Activation faults, also referred to as data faults, involve bit-flips in the input or intermediate activations due to noisy interconnects, voltage dips, or electromagnetic (EM) interference. Weight faults, or model faults, refer to bit-flips in stored DNN parameters caused by memory corruption, transmission noise, or intentional attacker manipulation.

\begin{figure}[ht]
\centering
\includegraphics[width=0.9\linewidth]{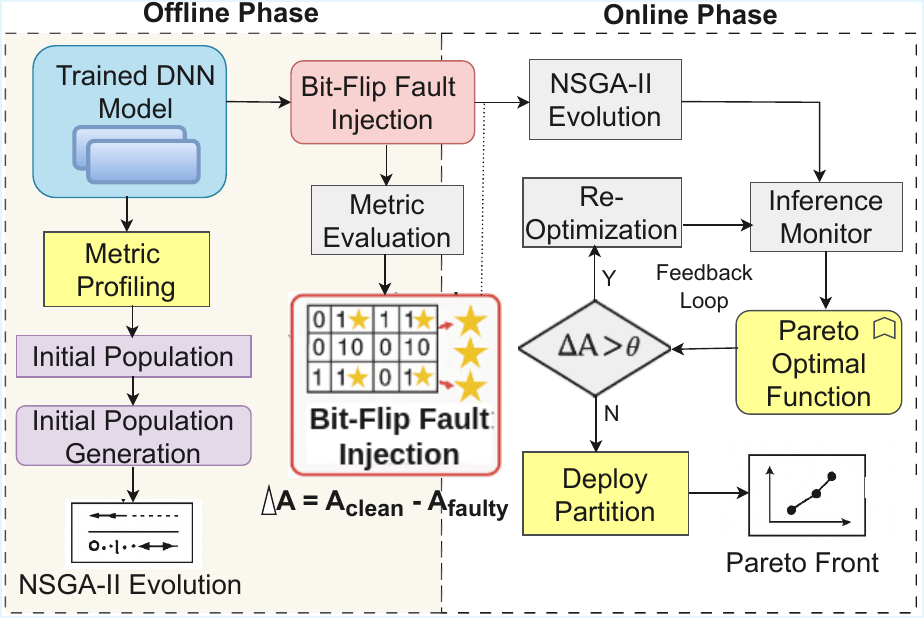}
\caption{Overview of our \emph{AFarePart} Framework}
\label{fig:framework}
\end{figure}


\section{Problem Definition}
\label{sec:prob-def}

Given a pretrained, quantized Deep Neural Network (DNN) consisting of $L$ layers, where the weights of  each layer are represented by $N_q$-bit signed fixed-point integers in 2’s complement format, the goal is to determine a fault-resilient partitioning of layers across a set of $D$ heterogeneous hardware accelerators (e.g., Eyeriss and SIMBA). Let $W_l$ denote the quantized weights of layer $l \in \{1, 2, \dots, L\}$, and $\hat{W}_l$ denote the faulty version of $W_l$ with bit-flips applied to the $b$ least significant bits (LSBs). Similarly, let $A_l$ denote the input activation tensor to layer $l$, and $\hat{A}_l$ its faulty version with LSB bit-flips. The layer-to-device mapping is represented by a function $P: \{1, 2, \dots, L\} \rightarrow \{0, 1, \dots, D-1\}$, where $P(l)$ assigns layer $l$ to device $d$.  

\textit{Definition: Accuracy Drop} - 
Let $(x, t)$ denote an input sample $x$ and its corresponding ground truth label $t$ from the dataset $\mathcal{D}$.
Let $Acc(\cdot, \cdot)$ denote the Top-1 classification accuracy metric, computed by comparing model predictions with ground-truth labels under a specified fault condition (fault-free or faulty).
The accuracy drop $\Delta Acc(P)$ for a given partition $P$ is defined as:
\begin{equation}
\Delta Acc(P) = Acc(f(x; W, A), t) - Acc(f(x; \hat{W}, \hat{A}), t)
\end{equation}
A larger $\Delta\mathrm{Acc}(P)$ indicates greater sensitivity to faults, while smaller values correspond to higher fault resilience.

The objective is to find a mapping $P$ that jointly minimizes inference latency, energy consumption, and accuracy drop under fault injection:
\begin{equation}
\min_{P}\; \big[\,\mathrm{Latency}(P),\; \mathrm{Energy}(P),\; \Delta\mathrm{Acc}(P)\,\big].
\end{equation}
The optimization is subject to: (i) a fault injection budget $N_b$, limiting the total number of bit-flips across the model; (ii) fault domain constraints, restricting faults to layers mapped to specific accelerators; and (iii) hardware resource constraints, as each device imposes its own compute and memory limits. NSGA-II is employed to discover Pareto-optimal partitions that simultaneously minimize all three objectives,  by efficient exploration of this constrained search space.

\section{\emph{AFarePart}: An Accuracy-aware Fault-resilient Partitioning of DNN}
\label{sec:framework}

We propose \emph{AFarePart}, a hybrid (offline–online) framework that enables fault-aware and resource-efficient deployment of Deep Neural Networks (DNNs) on heterogeneous edge platforms. As illustrated in Algorithm~\ref{alg:faultaware} and ~\autoref{fig:framework}, the framework operates in two coordinated phases: Offline Optimization and Online Deployment.

\subsection{Offline Phase - Multi-Objective Partitioning}
Given a pretrained and quantized DNN (e.g. a CNN, ResNet18 in ONNX format), the model is first profiled to gather layer-wise latency and energy estimates on multiple accelerators (e.g., Eyeriss, SIMBA). Fault injection is then performed by probabilistically flipping bits in activations and weights to simulate realistic soft errors, allowing evaluation of accuracy degradation. Using this data, an initial population of random layer-to-device mappings is evolved using the NSGA-II multi-objective evolutionary algorithm to minimize three key objectives: latency ($L$), energy ($E$), and accuracy loss ($\Delta A = A_{\text{clean}} - A_{\text{faulty}}$). Each candidate partition is evaluated under both fault-free and fault-injected conditions. Genetic operators such as selection, crossover, and mutation refine the partition space over \( G \) generations, producing a Pareto front of solutions that balance performance, efficiency and fault tolerance.

\begin{algorithm}[htbp]
\DontPrintSemicolon
\caption{Fault-Aware Partitioning with Dynamic Reconfiguration}
\label{alg:faultaware}

\KwIn{DNN model $M$, fault model $F$, profiling data $P$, population size $N$, generations $G$, accuracy threshold $\theta$}
\KwOut{Resilient partition $P^*$ minimizing latency, energy, and accuracy drop}

\tcc*[h]{Offline Phase: Multi-objective Optimization via NSGA-II}

Load model $M$ and profiling data $P$ \;
Generate initial population $\mathcal{P} = \{P_1, P_2, ..., P_N\}$ \;

\For{$g \gets 1$ \KwTo $G$}{
    \ForEach{$P_i \in \mathcal{P}$}{
        $A_{\text{clean}} \gets$ EvaluateAccuracy($M$, $P_i$, $F = \text{False}$) \;
        $A_{\text{faulty}} \gets$ EvaluateAccuracy($M$, $P_i$, $F$) \;
        $\Delta A(P_i) \gets A_{\text{clean}} - A_{\text{faulty}}$ \;
        $L(P_i) \gets$ EstimateLatency($P_i$, $P$) \;
        $E(P_i) \gets$ EstimateEnergy($P_i$, $P$) \;
        $F(P_i) \gets \{L(P_i), E(P_i), \Delta A(P_i)\}$ \tcp*{$F(P_i)$ is combined objective vector}
    }
    Apply NSGA-II operators: selection, crossover, mutation \;
}

Return Pareto front $\mathcal{P}^*$ \;

\tcc*[h]{Online Phase: Dynamic Accuracy-Aware Repartitioning}

Deploy $P^* \in \mathcal{P}^*$ \;
\While{inference is running}{
    $A_{\text{faulty}} \gets$ Evaluate($P^*$, $F$) \;
    \If{$A_{\text{clean}} - A_{\text{faulty}} > \theta$}{
        $P' \gets$ RunNSGAIIWithCurrentStats() \;
        $P^* \gets P'$ \tcp*{Re-optimize and update partition using current latency, energy, and accuracy degradation as feedback}
    }
}
\Return{$P^*$}
\end{algorithm}

\subsection{Online Phase – Dynamic Reconfiguration for Resilience}
During inference, the system operates with the most robust partition $P^*$ selected from the offline Pareto front, ensuring an initial balance between latency, energy and fault resilience. If the observed accuracy degradation exceeds a user-defined threshold \(\Delta A(P^*)>\theta\), the system triggers dynamic repartitioning. As shown in Algorithm~\ref{alg:faultaware}, this is handled by \texttt{RunNSGAIIWithCurrentStats}(), which re-invokes the NSGA-II optimization using current runtime profiling data (latency, energy, and accuracy drop) to adaptively generate an updated and more resilient partition \(P'\). Unlike the offline phase in which static pre-collected benchmarks are used, the online phase leverages real-time estimates of layer-wise latency, energy consumption and fault-induced accuracy degradation. These updated metrics reflect the current operational state of the hardware accelerators and the fault environment. 

The NSGA-II optimizer re-evaluates the search space of layer-to-device mappings under these updated conditions and produces a new set of Pareto-optimal candidate partitions. From these, a new partition \(P'\) is selected based on its ability to reduce the observed accuracy drop while keeping latency and energy within acceptable limits. This re-optimized partition \(P'\) replaces the previous configuration $P^*$, enabling the system to continue inference with improved fault resilience.
By dynamically remapping sensitive layers to more reliable devices or adjusting the partition to avoid overloading fault-prone accelerators, the system maintains robust and efficient inference throughout deployment.


\subsection{Fault Injection Methodology}
We adopt a probabilistic bit-flip model, applied to the least significant bits (LSBs) of fixed-point tensors as done in the prior works~\cite{Bit-Flip-Crushing, sifi}. Each bit in the vulnerable LSB range has an independent probability of flipping, emulating transient soft faults with configurable fault rates and bit-level corruption. The bit-flip fault injection process is depicted in Algorithm~\ref{alg:bitflip}.
Faults are injected during each fitness evaluation in the NSGA-II loop using two strategies: layer-wise fault sweeping, where faults are introduced in one layer at a time, and platform-specific injection, where faults are applied to layers mapped to a particular accelerator. This enables both uniform and targeted fault simulations aligned with the current partitioning.

\begin{algorithm}[htbp]
\DontPrintSemicolon
\caption{Bit-Flip Fault Injection Methodology}
\label{alg:bitflip}

\KwIn{Tensor $T$,fault rate $\text{FR}$,vulnerable LSB count $b$}
\KwOut{Fault-injected tensor $T_f$}

$T_f \gets \text{copy}(T)$ \tcp*{Clone the input tensor}

\ForEach{element $x \in T_f$}{
    \For{$i \gets 0$ \KwTo $b-1$}{
        \If{$\text{rand()} < \text{FR}$}{
            $x \gets x \oplus (1 \ll i)$ \tcp*{Flip bit $i$}
        }
    }
}
\Return $T_f$
\end{algorithm}

\section{Experimental Results}
\label{sec:results}

\subsection{Experimental Setup}

The proposed DNN partitioner \textit{AFarePart} is implemented as Python modules. We have also developed a \textit{fault-unaware} base model using NSGA-II with latency and energy as optimization metrics, for the sake of comparison. We run both the fault-aware and fault-unaware models for $60$ generations with a population size of $60$, using an accuracy drop threshold of  $1\%$ to trigger dynamic repartitioning. The experimental data for CNNParted~\cite{cnnparted} are obtained by running it at our end. 

Our experiments use real execution profiles measured on Eyeriss~\cite{chen2019eyeriss} and standard image classification models (e.g., \textit{AlexNet}, \textit{Squeezenet} and \textit{ResNet18}). 
The simulation is performed in Timeloop~\cite{timeloop} for latency and Accelergy~\cite{timeloop} for energy. For the SIMBA accelerator~\cite{symba}, analytical profiling is used. The models are originally trained on the ImageNet dataset using pre-trained weights. For fault-injection analysis, we use \textit{Tiny-ImageNet} dataset. Although our framework supports layerwise fault injection, here we evaluated the CNNs for faults in the earlier convolution layers. 

\subsection{Fault Injection}
We assume transient faults affecting the least significant bits (LSBs) of the input activations and weights. Faults are injected at configurable rates (e.g., $10\%$ to $40\%$), and corruption is applied to the lower 4 bits of 16-bit fixed-point representations. Both independent and simultaneous fault injections in input and weight tensors are analyzed.

\noindent An example configuration for parameters are given as follows:

\begin{itemize}
  \item \texttt{precision}: 16-bit fixed-point
  \item \texttt{fault\_rates}: [2e-1, 2e-1] (for activations and weights, implies each bit has 20\% chance of being flipped)
  \item \texttt{faulty\_bits}: 4 (number of LSBs susceptible)
\end{itemize}

\begin{figure}[!t]
    \centering
    \resizebox{\columnwidth}{!}{%
    \begin{tikzpicture}
            \begin{axis}[
                ybar,
                bar width=4pt,
                height=4cm,
                enlarge x limits=0.2,
                ylabel={\scriptsize Acc. (\%)},
                ylabel style={font=\scriptsize},
                symbolic x coords={AlexNet, SqueezeNet, ResNet18},
                xtick=data,
                xtick style={draw=none},
                xticklabel style={font=\tiny},
                ymin=50, ymax=94,
                nodes near coords,
                nodes near coords align={vertical},
                nodes near coords style={font=\tiny},
                legend style={
    at={(0.5,-0.28)},
    anchor=north,
    legend columns=3,
    font=\tiny,
    /tikz/every even column/.append style={column sep=1pt}, 
    /tikz/column 2/.style={column sep=1pt},                  
    inner sep=1pt,                                           
    row sep=-1pt                                             
}
,
                tick label style={font=\scriptsize},
            ]
            \addplot+[style={fill=blue!30}] coordinates {(AlexNet,74.2) (SqueezeNet,67.7) (ResNet18,83.9)};
            \addplot+[style={fill=green!50}] coordinates {(AlexNet,72.0) (SqueezeNet,68.3) (ResNet18,82.1)};
            \addplot+[style={fill=red!50}] coordinates {(AlexNet,81.0) (SqueezeNet, 76.5) (ResNet18,88.4)};
            \legend{CNNParted, Fault-unaware, AFarePart}
            \end{axis}
        \end{tikzpicture}
        }
    \caption{Top-1 accuracy of CNNParted, Fault-unaware, and \emph{AFarePart} across three CNNs at fault rate $20\%$ in weights.}
    \label{fig:CNNaccuracy}
\end{figure}
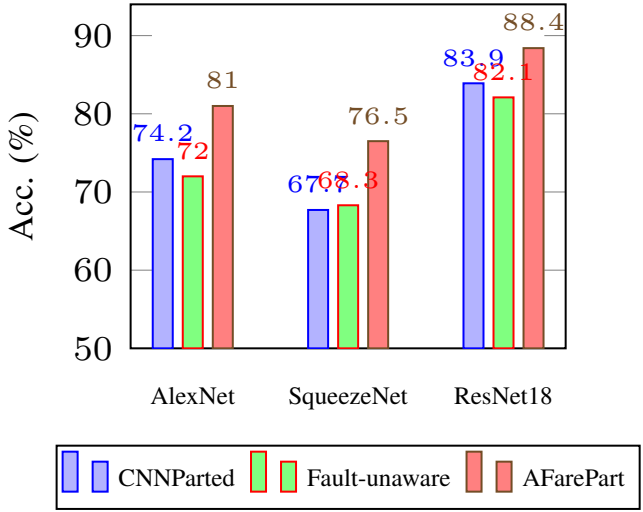

\begin{figure}[htbp]
\centering
 \resizebox{\linewidth}{!}{%
        \begin{tikzpicture}
            \begin{semilogxaxis}[
                width=\linewidth,
                height=5.5cm,
                xlabel={Fault Rate (log scale)},
                ylabel={Accuracy (\%)},
                legend pos=south west,
                grid=both,
                ymin=73, ymax=93,
                log ticks with fixed point,
                xtick={1e-1, 2e-1, 3e-1, 4e-1},
                xticklabels={$10\%$, $20\%$, $30\%$, $40\%$},
                ylabel near ticks,
                xlabel near ticks,
                tick label style={font=\small},
                legend style={font=\small}
            ]

            \addplot+[mark=*, thick, color=blue] coordinates {
                (1e-1, 87.9)
                (2e-1, 83.9)
                (3e-1, 78.6)
                (4e-1, 76.3)
            };
            \addlegendentry{CNNParted}

            \addplot+[mark=triangle*, thick, color=green] coordinates {
                (1e-1, 82.0)
                (2e-1, 82.1)
                (3e-1, 79.0)
                (4e-1, 77.1)
            };
            \addlegendentry{Fault-unaware}

            \addplot+[mark=square*, thick, color=red!60!black] coordinates {
                (1e-1, 89.9)
                (2e-1, 88.4)
                (3e-1, 86.0)
                (4e-1, 82.4)
            };
            \addlegendentry{AFarePart}

            \end{semilogxaxis}
        \end{tikzpicture}
        }

\caption{Accuracy vs. Fault Rate of different partitioning strategies for fault in weights for \textit{ResNet18}.}
\label{fig:acc_vs_fault}
\end{figure}
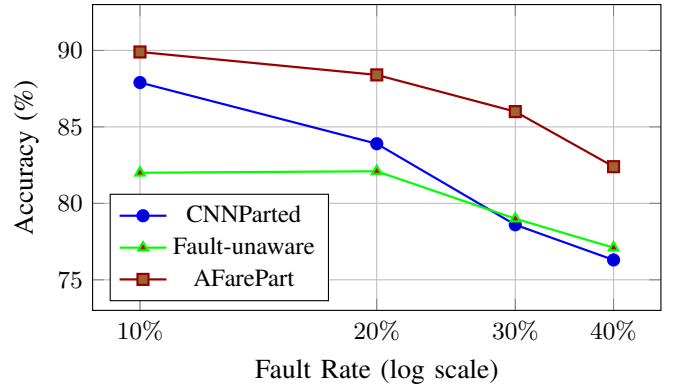

\subsection{Accuracy vs Fault Resilience Trade-off}
Figure~\ref{fig:CNNaccuracy} shows the top-1 accuracy under fault rate of $20\%$ for \textit{AlexNet}, \textit{Squeezenet} and \textit{ResNet18}. 
We observe that fault-aware partitioning by \emph{AFarePart} results in up to 9\% less accuracy degradation compared to baseline fault-unaware partitioning techniques. 
We also evaluate the impact of fault injection on partitioned CNNs under various fault rates. The results are summarized in Figure~\ref{fig:acc_vs_fault}. We observe that as the fault rate increases, the accuracy of the model drops significantly. 
However, in our proposed fault-aware partitioning framework, the accuracy drop is explicitly incorporated into the NSGA-II optimization process as one of the objective functions, thereby reducing accuracy loss and resulting in more robust partitions.

\begin{table*}[ht]
\setlength{\tabcolsep}{5pt}
\renewcommand{\arraystretch}{1.1}
\centering
\caption{ Comparison of Performance with $20\%$ Fault Rate across Fault Scenarios}
\begin{small}
\begin{tabular}{p{1.4cm}p{1.5cm}ccc|ccc|ccc}
\toprule
\multirow{2}{*}{\textbf{Model}} & \multirow{2}{*}{\textbf{Tool}} 
& \multicolumn{3}{c|}{\textbf{Weight Fault Only}} 
& \multicolumn{3}{c|}{\textbf{Input Fault Only}} 
& \multicolumn{3}{c}{\textbf{Input + Weight Fault}} \\
& & \textbf{Acc. (\%)} & \textbf{Lat. (ms)} & \textbf{Energy (mJ)} 
  & \textbf{Acc. (\%)} & \textbf{Lat. (ms)} & \textbf{Energy (mJ)} 
  & \textbf{Acc. (\%)} & \textbf{Lat. (ms)} & \textbf{Energy (mJ)} \\
\midrule
AlexNet 
& CNNParted     & 74.2 & 26.6 & 16.79 & 61.9 & 26.6 & 22.5 & 41.5 & 26.6 & 22.9 \\
& Flt-unware   & 72.0 & 21.0 & 13.2 & 61.5 & 21.0 & 14.2 & 43.2 & 21.0 & 15.8 \\
& \textbf{\textit{AFarePart}} 
               & \textbf{81.0} & 29.8 & 18.5 & \textbf{78.5} & 29.8 & 19.8 & \textbf{69.2} & 30.2 & 20.1 \\
\midrule
SqueezeNet 
& CNNParted     & 67.7 & 13.2 & 4.03 & 70.1 & 13.12 & 4.1 &38.2 & 13.2 & 4.3 \\
& Flt-unware   & 68.3 & 13.8 & 4.5 & 69.0 & 13.8 & 4.6 & 40.1 & 13.8 & 4.9 \\
& \textbf{\textit{AFarePart}} 
               & \textbf{76.5} & 15.0 & 5.2 & \textbf{74.2} & 15.0 & 5.5 & \textbf{61.8} & 15.5 & 6.0 \\
\midrule
ResNet18 
& CNNParted     & 83.9 & 65.9 & 19.1 & 77.0 &  65.9 & 18.1 & 70.5 & 65.9 & 19.6 \\
& Flt-unware   & 82.1 & 61.5 & 18.0 & 77.5 & 61.5 & 18.2 & 72.0 & 61.5 & 18.5 \\
& \textbf{\textit{AFarePart}} 
               & \textbf{88.4} & 67.0 & 20.4 & \textbf{85.6} & 67.0 & 20.5 & \textbf{81.2} & 67.5 & 21.0 \\
\bottomrule
\end{tabular}
\end{small}
\label{tab:comp_all_faults}
\end{table*}

\subsection{Latency and Energy Overhead Analysis}

As shown in Table~\ref{tab:comp_all_faults}, the proposed approach consistently achieves higher Top-1 accuracy under weight-only, input-only, and combined input+weight fault scenarios compared to CNNParted and fault-unaware baseline. We observe up to 27.7\% higher accuracy compared to CNNParted under the combined input and weight fault scenario, demonstrating strong resilience in the most challenging fault environments. This enhanced robustness comes with a modest performance trade-off: latency increases by approximately 9.7\%, and energy overhead is limited to around 4.3\% on average, relative to CNNParted. These increases are due to the strategic placement of critical layers on more reliable accelerators and improved layer-to-device mapping choices that prioritize fault resilience. 

Although both CNNParted and our fault-unaware baseline are fault-agnostic, their performance differs slightly due to differences in optimization heuristics and objective weighting. CNNParted, with its emphasis on aggressive latency and energy minimization, may inadvertently assign critical layers to more error-prone accelerators. In contrast, our baseline employs alternative partitioning strategies that, in some cases, result in more resilient layer mappings despite the absence of explicit fault-awareness.


\subsection{Discussions}
\label{sec:disc}

In this work we focus on faults targeting the least significant bits (LSBs), as most significant bit (MSB) errors cause drastic accuracy loss and require mitigation beyond partitioning. While LSB flips seem less severe, they accumulate through layers, leading to significant accuracy degradation. 
\textit{AFarePart} dynamically adjusts partitioning decisions based on insights derived from fault simulation feedback. Unlike CNNParted, \textit{AFarePart} currently excludes link latency and link energy to keep optimization focused on mitigating accuracy degradation. However, these can be easily included and the results will be reported in the near future.

\section{Conclusion}
\label{sec:conclusion}
We propose a novel feedback-guided partitioning framework for DNNs that explicitly integrates fault resilience into the optimization process. By incorporating fault-awareness, our method enhances accuracy under fault conditions while incurring minimal performance overhead, thus enabling reliable inference on heterogeneous and fault-prone System-on-Chip (SoC) platforms. To the best of our knowledge, this is the first work to embed fault injection-based analysis directly into a DNN partitioning methodology. This contribution addresses a critical gap in the existing literature and provides a practical foundation for deploying robust DNNs on resource-constrained hardware. As future work, we aim to extend this framework to platforms such as FPGAs, enabling real-time validation of fault-aware partitioning through partial reconfiguration or bitstream-level fault injection.



\bibliographystyle{IEEEtran}
\bibliography{sample}

\end{document}